\documentclass[english]{article}
\usepackage[T1]{fontenc}
\usepackage[latin9]{inputenc}
\usepackage{geometry}
\usepackage{float}
\usepackage{graphicx}
\usepackage{setspace}
\makeatletter
\date{}

\makeatother

\usepackage{babel}

\begin{document}

\title{Authenticated Semi-quantum Direct Communication Protocols using Bell
States}

\author{Yi-Ping Luo and Tzonelih Hwang%
\thanks{Corresponding Author, email: hwangtl@ismail.csie.ncku.edu.tw%
} }
\maketitle
\begin{abstract}
This study presents the first two authenticated semi-quantum direct
communication (ASQDC) protocols without using any classical channel.
By pre-sharing the master secret key between two communicants, a sender
with advanced quantum devices can transmit a secret message to a receiver
who can only perform classical operations without any information
leakage. The receiver is then capable of verifying the message up
to the single qubit level, i.e., a one-qubit modification of the transmitted
quantum sequence can be detected with a probability close to 1. Moreover,
the proposed protocols are resistant to several well-known attacks.\end{abstract}
\begin{description}
\item [{Keywords:}] Authentication; Authenticated semi-quantum communication;
Bell states; Quantum communication; Quantum cryptography; Semi-quantum
communication.
\end{description}

\section{Introduction}

Authentication, which is a process used for guaranteeing the integrity
and origin of a transmitted message, is an important topic in information
security. Due to authentication of the message, the receiver (the
verifier) can determine whether or not the receiver is communicating
with the alleged participant (the sender). Moreover, authentication
can be used to verify the integrity of the received message without
any modification. The feature of authentication is a very important
requirement in various quantum cryptographic environments, such as
quantum key distribution (QKD) protocols, quantum secure direct communications
(QSDC), deterministic secure quantum communications (DSQC), quantum
dialogue (QD), quantum private comparison (QPC), and quantum secret
sharing (QSS). To simplify the design, however, the majority of the
abovementioned environments focus on the following two methods of
providing secrecy as well as authentication.
\begin{enumerate}
\item An authenticated classical channel (i.e., transmitted information
that can be eavesdropped but not modified) is assumed to be available
for providing authentication, which can be further used for detecting
eavesdropping. Accordingly, both the information integrity and originality
can be guaranteed. In practice, however, if two communicants want
to communicate with each other, the QKD or QSDC protocol must be performed
whenever a communication session is initiated. That is, both communicants
must be in an environment where an authenticated classical channel
is available, which could be a restriction for some applications.
For example, a traveling mobile user will have difficulty maintaining
an authenticated classical channel with low-power mobile devices.
\item All participants are required to have quantum capabilities. That is,
the protocol requires that every participant has access to quantum
memory and has the ability to prepare/measure arbitrary quantum states
and to perform operations. However, not all participants can afford
such expensive quantum resources and operations for various applications.
In this case, it will be difficult to apply these protocols in practical
environments.
\end{enumerate}
To resolve these issues, Yu et al. (2014) proposed authenticated semi-quantum
key distribution (ASQKD) protocols \cite{Yu2014}. In these protocols,
by pre-sharing a master secret key between two communicants, a sender
with advanced quantum devices can transmit a working key to a receiver,
who can only perform classical operations, without requiring an authenticated
classical channel. According to the definition in \cite{Boy2007,Boy2009},
the term \textquotedblleft{}\textbf{semi-quantum}\textquotedblright{}
implies that the sender, Alice, is a powerful quantum communicant,
whereas the receiver, Bob, has only classical capabilities. More precisely,
the sender (Alice) has the ability to perform following operations:
(1) prepare any quantum state such as single photons and Bell states,
(2) perform any quantum measurement such as Bell measurement and multi-qubit
joint measurement, and (3) store qubits in a quantum memory. Conversely,
the classical receiver (Bob) is restricted to performing the following
operations over the quantum channel: (1) prepare new qubits in the
classical basis $\left\{ \left|0\right\rangle ,\left|1\right\rangle \right\} $
(i.e., the Z basis), (2) measure qubits in the classical basis, (3)
reorder the qubits via different delay lines, and (4) send or reflect
the qubits without disturbance. Because the classical basis only considers
the qubits $\left|0\right\rangle $ and $\left|1\right\rangle $,
other quantum superpositions of single photons are not considered.
Hence, the operations performed by Bob are equivalent to traditional
$\left\{ 0,1\right\} $ computation. Following Boyer et al., Yu et
al. also proposed two types of ASQKD protocols, namely randomization-based
ASQKD and measure-resend ASQKD. The difference between these two protocols
lies in the capability of the classical Bob. In the randomization-based
ASQKD protocol, classical Bob is limited to performing operations
(2), (3), and (4), whereas in the measure-resend ASQKD protocol, classical
Bob is limited to performing operations (1), (2), and (4). Because
the ASQKD protocol allows a classical Bob to be a receiver and does
not require an authenticated classical channel, an authenticated semi-quantum
protocol can reduce not only the computational burden of the communicants
but also the cost of the quantum hardware devices in practical implementations.

In this paper, we propose authenticated semi-quantum direct communication
(ASQDC) protocols using Bell states. To the best of our knowledge,
there is no existing ASQDC protocol that enables the quantum sender
to directly send a secret message to the classical receiver without
any information leakage. Furthermore, the proposed ASQDC protocols
have the following features:
\begin{enumerate}
\item The protocols do not require any classical channel.
\item The pre-shared secret key between two communicants can be reused.
\item The protocols can satisfy the requirements of a quantum direct communication
protocol, which was defined by Deng et al. \cite{Den2003}. First,
the secret messages should be directly read out by the legitimate
user Bob when he receives the quantum states, and no additional classical
information is needed after the qubit transmission. Second, the secret
messages, which have been previously encoded with quantum states,
should not leak even though an eavesdropper may access the channel.
\item The security of the proposed ASQDC protocols is guaranteed by quantum
mechanics, i.e., by the uncertainty of quantum measurement and the
no-cloning theory.
\item The protocols can resist impersonation attacks, intercept-and-resend
attacks, modification attacks, and other well-known attacks.
\end{enumerate}
Moreover, the proposed ASQDC protocols, together with the semi-QKD
or QKD protocol, are more effective than merely performing the QSDC
protocol to solve the abovementioned scenario. Hence, the proposed
ASQDC protocols will be more practical in various environments such
as sensor communications, in which a sensor with limited capability
wants to collect or receive secret information based on the security
provided by quantum mechanics.

The remainder of this paper is organized as follows. Section 2 presents
our ASQDC protocols using Bell states, and Section 3 provides security
analyses of the proposed protocols. Finally, our conclusions are given
in Section 4.

\section{Proposed ASQDC Protocols}

This section presents two ASQDC protocols, which enable a quantum
sender, Alice, to send an $\frac{n}{8}$-bit secret message $m$ to
a classical receiver, Bob. In Section 2.1, the randomization-based
protocol is proposed. After that, the measure-resend one is given.

\subsection{Randomization-based ASQDC protocol}

Let us assume that Alice and Bob pre-shared two secret keys $K_{1}$
and $K_{2}$, where $K_{1}\in\left\{ 0,1\right\} ^{n}$ and $K_{2}\in\left\{ 0,1\right\} ^{\frac{n}{2}}$.
Besides, the quantum channels here are assumed to be noiseless and
lossless. The procedure of the randomization-based ASQDC is described
in the following steps (see also Figure 1):
\begin{description}
\item [{Step\ 1.}] Alice calculates $M=m||h\left(m\right)$, where `||'
denotes concatenation and $h\left(\right)$ is a one-way hash function
\cite{FIPS1995,Pre1997} to produce an $\frac{n}{8}$-bit checking
value of $m$. After that, Alice generates a sequence of Bell states,
$S=\left\{ s_{1},s_{2},...,s_{\frac{n}{4}}\right\} $, based on $M$,
where $s_{i}=\left\{ q_{1}^{i},q_{2}^{i}\right\} $ for $i=1,2,...,\frac{n}{4}$.
That is, if the $i$th bit of $M$ is zero, i.e., $M^{i}=0$, Alice
produces $s_{i}$ in $\left|\Phi^{+}\right\rangle =\frac{1}{\sqrt{2}}\left(\left|00\right\rangle +\left|11\right\rangle \right)$.
Otherwise, Alice produces $\left|\Psi^{-}\right\rangle =\frac{1}{\sqrt{2}}\left(\left|01\right\rangle +\left|10\right\rangle \right)$.
Then, Alice generates the checking state $C=\left\{ c_{1},c_{2},...,c_{\frac{n}{2}}\right\} $
randomly in the states of $\left|\Phi^{+}\right\rangle $ and $\left|\Psi^{-}\right\rangle $
whose initial state is denoted as $IS_{C}$, where $c_{j}=\left\{ qc_{1}^{j},qc_{2}^{j}\right\} $
for $j=1,2,...,\frac{n}{2}$. After that, she divides these $\frac{n}{2}$
Bell states into two ordered sequences, $C_{A}=\left\{ qc_{1}^{j}\right\} $
and $C_{B}=\left\{ qc_{2}^{j}\right\} $. Now, she reorders the quantum
sequences $S$ and $C_{B}$ together according to the secret key $K_{1}$
to obtain the new quantum sequence $Q$. After the above preparation,
Alice retains the sequence $C_{A}$ and sends the sequence $Q$ to
Bob.
\item [{Step\ 2.}] When Bob receives the qubits in $Q$, he puts every
qubit into the delay line device whose traveling time is long enough
to wait for the last qubit enters so that he can get the ordered sequence
$S'$ and $C_{B}'$ according to $K_{1}$. After that, he performs
a Z-basis measurement on each qubit in $S'$ and obtains the measurement
result $MR_{B}$. Then, Bob can calculate $\left(M'\right)^{i}=MR_{B}^{2i-1}\oplus MR_{B}^{2i}$
to derive $M'=m'||h\left(m\right)'$. That is, if $MR_{B}=00$ (11),
then $M'=0\oplus0=0$ ($1\oplus1=0$). If $MR_{B}=01$ (10), then
$M'=0\oplus1=1$ ($1\oplus0=1$). Bob calculates $h\left(m'\right)$
and compares it with the received $h\left(m\right)'$. If $h\left(m'\right)=h\left(m\right)'$,
Bob believes that the message $m'$ is indeed sent from Alice without
any disturbance. Otherwise, Alice and Bob will terminate the protocol
and start it again.
\item [{Step\ 3.}] Bob reorders the qubits $C_{B}'$ based on the secret
key $K_{2}$ to obtain $C_{B}''$ and reflects $C_{B}''$ back to
Alice via different delay lines. 
\item [{Step\ 4.}] Upon receiving $C_{B}''$, Alice can recover the reflected
qubits in the correct order based on $K_{2}$ to obtain the ordered
sequence $C_{B}'=\left\{ \left(qc_{2}'\right)^{j}\right\} $. Then
Alice performs Bell measurement on $\left\{ qc_{1}^{j},\left(qc_{2}'\right)^{j}\right\} $
for $j=1,2,...,\frac{n}{2}$ to obtain $IS_{C}'$ and then check whether
each corresponding set of two qubits in $IS_{C}'$ is consistent with
the states she generated in Step 1, $IS_{C}$. If the transmission
between Alice and Bob is secure, then it means Alice has successfully
transmitted the secret message to Bob. %
\begin{figure}[H]
\includegraphics[scale=0.55]{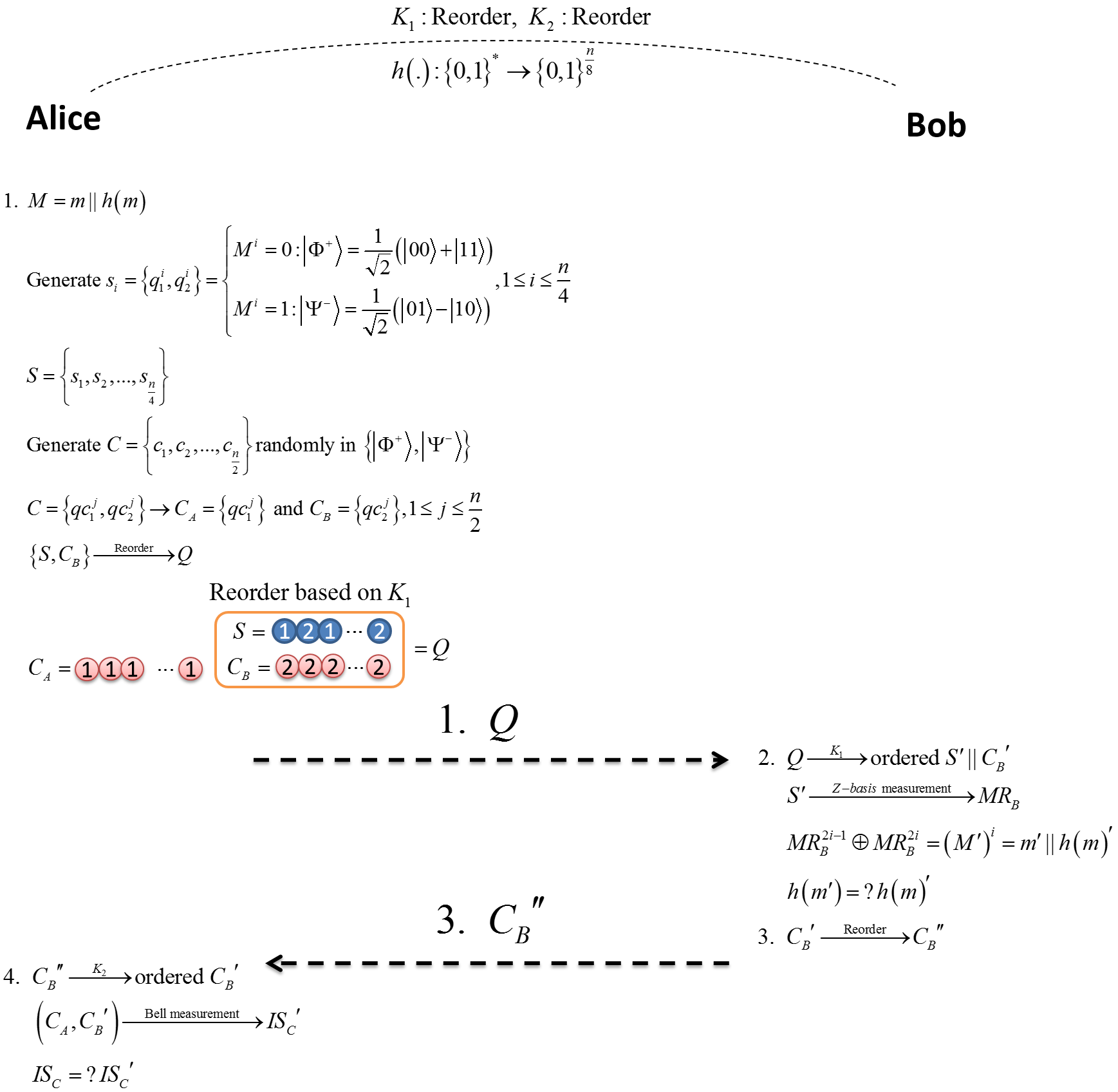}\caption{The proposed randomization-based ASQDC protocol}

\end{figure}

\end{description}

\subsection{Measure-resend ASQDC protocol}

Here, a measure-resend ASQDC protocol, which modifies the operations
that Bob is allowed to perform in the randomization-based ASQKD described
in section 2.1, is as follows (see also Figure 2). The modified steps
({*}) are listed in detail, as follows. The others are the same as
those described in section 2.1. In this case, we assume Alice and
Bob pre-share a secret key $K_{1}$, where $K_{1}\in\left\{ 0,1\right\} ^{n}$.
Besides, the quantum channels here are assumed to be noiseless and
lossless. 
\begin{description}
\item [{Step\ 2{*}.}] Based on the secret key $K_{1}$, Bob decides to
perform either SHARE or CHECK on each received qubit. In the SHARE
mode, Bob measures the received qubit using the Z basis to obtain
the measurement result $MR_{B}$ and returns a qubit of the same state
to Alice. However, in the CHECK mode, Bob reflects the qubit without
any disturbance back to Alice. Let assume the returned quantum sequence
is $Q'$.
\item [{Step\ 3{*}.}] Bob recovers $MR_{B}$ to the ordered sequence $MR_{B}'$
based on $K_{1}$. After that, he calculate $\left(M'\right)^{i}=\left(MR_{B}'\right)^{2i-1}\oplus\left(MR_{B}'\right)^{2i}$
to derive $M'=m'||h\left(m\right)'$. That is, if $MR_{B}=00$ (11),
then $M'=0\oplus0=0$ ($1\oplus1=0$). If $MR_{B}=01$ (10), then
$M'=0\oplus1=1$ ($1\oplus0=1$). Then, Bob calculates $h\left(m'\right)$
and compares it with the received $h\left(m\right)'$. If $h\left(m'\right)=h\left(m\right)'$,
Bob believes that the message $m'$ is indeed sent from Alice without
any disturbance. Otherwise, Alice and Bob will terminate the protocol
and start it again.
\item [{Step\ 4{*}.}] Upon receiving $Q'$, Alice can recover $Q'$ based
on $K_{1}$ to obtain the ordered sequence $Q''=S''||C_{B}''$. After
that, Alice performs Bell measurement on $\left\{ qc_{1}^{j},\left(qc_{2}''\right)^{j}\right\} $
for $j=1,2,...,\frac{n}{2}$ to check whether each corresponding set
of two qubits is consistent with the states she generated in Step
1. If there is no eavesdropper, Alice performs Bell measurement on
$s_{i}''=\left\{ \left(q''_{1}\right)^{i},\left(q''_{2}\right)^{i}\right\} $
for $i=1,2,...,\frac{n}{4}$. If the message is 0 (1), i.e., the initial
state is $\left|\Phi^{+}\right\rangle $ ($\left|\Psi^{-}\right\rangle $),
then the measurement result, $M_{B}$, is one of $\left\{ \left|\Phi^{+}\right\rangle ,\left|\Phi^{-}\right\rangle \right\} $
($\left\{ \left|\Psi^{+}\right\rangle ,\left|\Psi^{-}\right\rangle \right\} $).
If the measurement results are all in the same as their initial states
(i.e., $\left|\Phi^{+}\right\rangle $ or $\left|\Psi^{-}\right\rangle $),
then it indicates a reflecting attack, and hence, Alice and Bob will
terminate the protocol and start it again.%
\begin{figure}[H]

\includegraphics[scale=0.55]{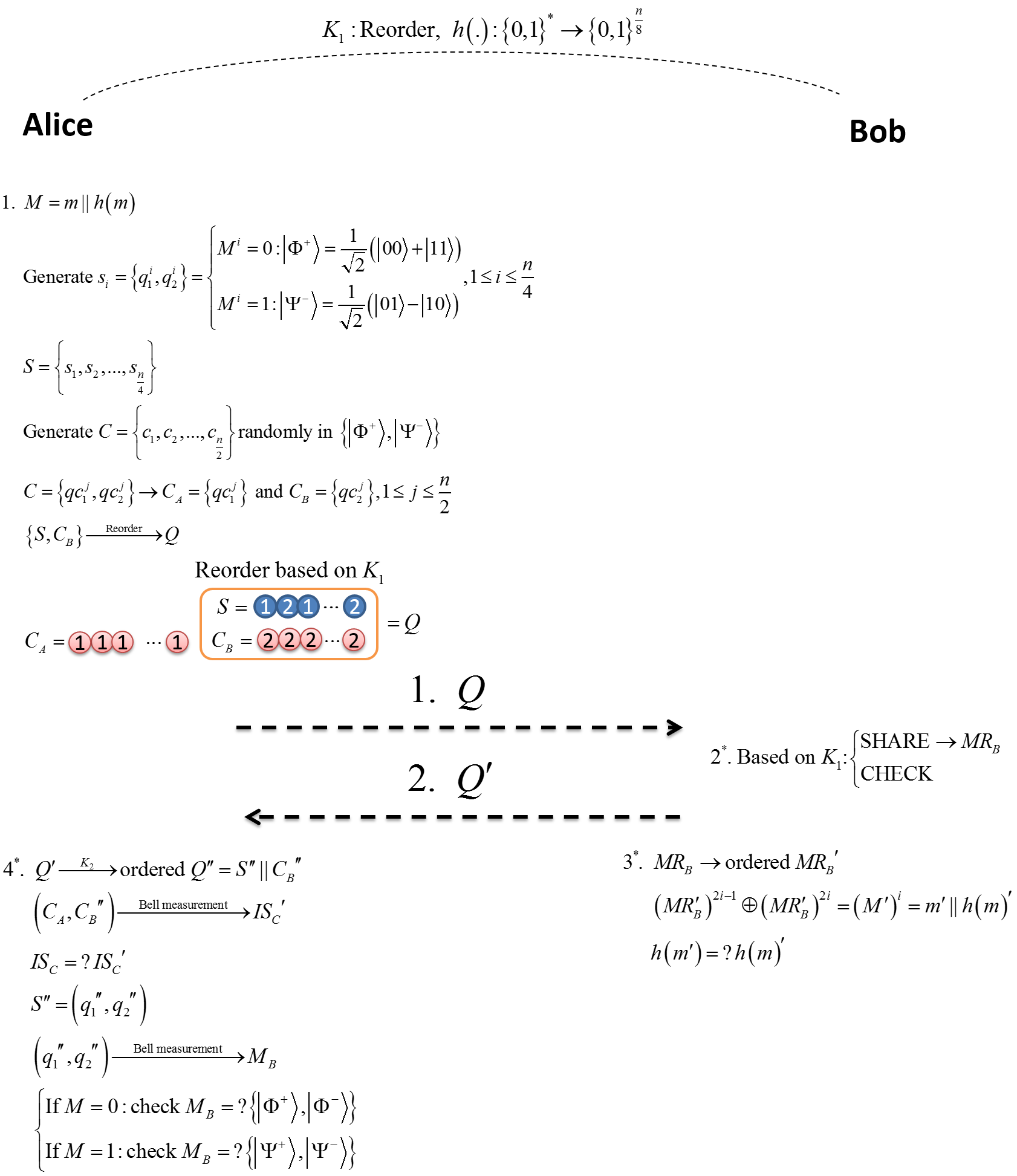}\caption{The proposed measure-resend ASQDC protocol}

\end{figure}

\end{description}
Both ASQDC protocols use the entanglement correlation of the Bell
state to detect the presence of eavesdroppers. The only difference
between these two protocols (the randomization-based ASQDC and the
measure-resend ASQDC) is in the type of operations that Bob is allowed
to perform in Step 2 and Step 2{*}. In the proposed ASQDC protocols,
a one-way hash function is used in the background for eavesdropping
check and for verifying the integrity of a secret message. Based on
the property of a one-way hash function, one-bit error in the input
(i.e., a transmitted message) will cause significant changes in the
output (i.e., a hashed value) and can be detected in Step 2 and Step
3{*}. This property is very useful in checking message integrity if
the quantum channel is reliable or ideal. In reality, however, some
states of the transmitted qubits may be changed due to the unexpected
interference of the optical fiber or due to the environment. These
changes of the transmitted qubits caused by noises will be detected
as an eavesdropping by the proposed protocol. In this situation, quantum
error correction codes \cite{Cal1996,Cal1997,Eke1996,Got1996,Kni2000,Laf1996,Sho1995}
can be applied after the hash function to solve this problem. As a
result, small errors can be corrected by the introduced quantum error
correction code and the majority errors due to malicious users can
be detected by the one-way hash function. By combining quantum error
correction code and a one-way hash function with quantum mechanics,
the proposed protocol can provide data privacy as well as message
integrity on the noisy quantum channel.

In the proposed ASQDC protocols, the pre-shared secret keys are used
for user authentication and message authentication. However, it should
be noted that the secret keys can be reused if no eavesdropper is
detected. Consequently, the communicants do not have to renew the
secret keys, which is a tedious work, after completing a protocol
execution. Only when a failure occurs in the eavesdropping check or
when the secret keys are used for a long period of time does the new
secret keys have to be shared again between Alice and Bob.

\section{Security Analyses}

In this section, three well-known attacks, i.e., the impersonation
attack, the intercept-and-resend attack, and the modification attack,
are analyzed respectively. It should be noted that only the security
of the randomization-based ASQDC protocol is analyzed in detail. As
for the security of the measure-resend ASQDC protocol, the same analysis
can be performed.

\subsection{Security against impersonation attack}

An attacker, Eve, may try to impersonate Alice to send a forged message
to Bob. Without knowing the pre-shared key $K_{1}$, however, Eve
will be caught by Bob with a very high probability. In the randomization-based
ASQDC protocol, suppose Eve generates a sequence of qubits, $Q_{E}$,
and sends them to Bob in Step 1. If Eve can pass the eavesdropping
check in Step 2, then she is able to successfully impersonate Alice
to send a forged message to Bob. However, without knowing the pre-shared
key $K_{1}$, Eve cannot perform the correct reorder operation on
$Q_{E}$ and eventually the comparison in Step 2 will be failed. Since
one-bit error in the input (i.e., a transmitted message) will cause
significant changes in the output (i.e., a hashed value), the probability
for Eve to be detected in the randomization-based ASQDC protocol is
close to 1.

On the other hand, Eve may try to impersonate Bob to communicate with
Alice. In the randomization-based ASQDC protocol, Eve may intercept
the sequence $Q$ sent from Alice to Bob in Step 1. Since Eve does
not know the secret key $K_{1}$ and $K_{2}$ hence she does not know
how to choose the reflecting qubits in $Q$ and does not know how
to perform the reorder operation on the chosen qubits, respectively.
In this case, she will randomly choose some qubits in $Q$ and randomly
reorders the chosen qubits and sends them to Alice in Step 3. If,
however, Eve reflects the wrong qubits with the wrong order back to
Alice, then Eve can successfully pass the verification process of
Alice with a probability of $\frac{1}{4}$ for each qubit. For example,
if the initial state is $\left|\Phi^{+}\right\rangle $ ($\left|\Psi^{-}\right\rangle $),
then Alice performs the Bell measurement on the wrong qubit to obtain
the measurement result $\left|\Phi^{+}\right\rangle $ ($\left|\Psi^{-}\right\rangle $)
with a probability of $\frac{1}{4}$ because she will randomly obtain
one of the four measurement results from $\left\{ \left|\Phi^{+}\right\rangle ,\left|\Phi^{-}\right\rangle ,\left|\Psi^{+}\right\rangle ,\left|\Psi^{-}\right\rangle \right\} $.
As a result, Eve has a probability of $\frac{5}{8}$ $\left(=\frac{1}{2}+\frac{1}{2}\times\frac{1}{4}\right)$
to pass the verification for each qubit. Hence, the probability for
Eve to be detected in the randomization-based ASQDC protocol is $1-\left(\frac{5}{8}\right)^{\frac{n}{2}}$.
The detection probability would converge to 1 when $n$ is large enough.

\subsection{Security against intercept-and-resend attack}

Eve may launch the intercept-and-resend attack in hope that she can
get the useful information about the secret message $M$ without being
detected. In this attack, Eve intercepts the sequence $Q$ in Step
1 and measures it with $Z$ basis $(\left\{ \left|0\right\rangle ,\left|1\right\rangle \right\} )$.
After that, she generates the same states based on her measurement
results and sends them to Bob. However, each secret message in $M$
is encoded into a Bell state $\left|\Phi^{+}\right\rangle =\frac{1}{\sqrt{2}}\left(\left|00\right\rangle +\left|11\right\rangle \right)$
(if $M^{i}=0$) or $\left|\Psi^{-}\right\rangle =\frac{1}{\sqrt{2}}\left(\left|01\right\rangle +\left|10\right\rangle \right)$
(if $M^{i}=1$) and then reordered with the checking states $C_{B}$
based on $K_{1}$. Without knowing the secret key $K_{1}$, Eve cannot
identify which qubit belongs to $S$ and which qubit belongs to $C_{B}$.
Therefore, Eve cannot recover the measurement results to the correct
order and hence cannot calculate the secret message of Alice. 

Besides, any arbitrary measurement on $C_{B}$ would destroy the entanglement
of a Bell state and eventually will be detected by Alice in Step 4
with a probability of $\frac{1}{2}$ for each Bell state in $C_{B}$.
For example, if the initial state of the Bell state is $\left|\Phi^{+}\right\rangle $,
after Eve performs the intercept-and-resend attack, the state of the
Bell state will collapse to $\left|00\right\rangle /$$\left|11\right\rangle $)
(since $\left|\Phi^{+}\right\rangle =\frac{1}{\sqrt{2}}\left(\left|00\right\rangle +\left|11\right\rangle \right)$).
In Step 4, Alice will perform the Bell measurement on $\left|00\right\rangle /$$\left|11\right\rangle $
and she will obtain the measurement result $\left|\Phi^{+}\right\rangle $
with a probability of $\frac{1}{2}$, since $\left|00\right\rangle =\frac{1}{\sqrt{2}}\left(\left|\Phi^{+}\right\rangle +\left|\Phi^{-}\right\rangle \right)$
and $\left|11\right\rangle =\frac{1}{\sqrt{2}}\left(\left|\Phi^{+}\right\rangle -\left|\Phi^{-}\right\rangle \right)$.
As a result, Eve has a probability of $\frac{1}{2}$ to pass the verification
for each Bell state. The probability for Eve to be detected in the
randomization-based ASQDC protocol is $1-\left(\frac{1}{2}\right)^{\frac{n}{2}}$.
The detection probability would converge to 1 when $n$ is large enough.

\subsection{Security against modification attack}

In the modification attack, Eve may try to modify one-bit message
of the transmitted qubits, $Q$, by using the unitary operation $i\sigma_{y}$
and make the receiver to obtain a wrong message without being detected.
The following two cases show that Eve will be detected by using the
checksum $h\left(m'\right)$ of the hash function or the entanglement
correlation of a Bell state as the integrity verification mechanism. 
\begin{enumerate}
\item If Eve performs the unitary operation $i\sigma_{y}$ on a qubit belong
to the sequence $S$ and then sends it to Bob. However, arbitrary
modification will lead to the wrong measurement result, and Bob can
detect the modification with 100\% probability in Step 2. This is
similar to the security analysis proposed in \cite{Yan2012,Yan2013_3,Hwa2014}:
if the 1-bit message is modified, then the computed checksum $h\left(m'\right)$
cannot be equal to the measured checksum, $h\left(m\right)'$, according
to the feature of a collision-free hash function.
\item If Eve performs the unitary operation $i\sigma_{y}$ on a qubit belong
to the sequence $C_{B}$ and then sends it to Bob. Then the the Bell
state $\left|\Phi^{+}\right\rangle $ ($\left|\Psi^{-}\right\rangle $)
will be changed to $\left|\Psi^{-}\right\rangle $ ($\left|\Phi^{+}\right\rangle $).
An arbitrary modification to a qubit, however, could lead to the wrong
measurement result and eventually would be detected by Alice. Hence,
Eve cannot pass the verification process of Alice because the measurement
result cannot be equal to the initial state.
\end{enumerate}
Therefore, the proposed ASQDC protocol is secure against the modification
attack to a single qubit level because Eve cannot modify the sequence
$Q$ without being detected.

\section{Conclusion}

In this paper, we propose two authenticated semi-quantum direct communication
(ASQDC) protocols without using classical channels. The first proposed
protocol is the randomization-based ASQDC protocol, and the other
protocol is based on the measure-resend ASQDC protocol. In both proposed
ASQDC protocols, a sender with advanced quantum devices can transmit
a secret message to a receiver, who can only perform classical operations,
without information leakage through the pre-shared secret keys. The
security analyses show that the proposed protocols are resistant to
the impersonation attack, the intercept-and-resend attack, and the
modification attack. However, we should note that, like all semi-quantum
schemes, the proposed protocols suffer from Trojan-horse attacks \cite{Cai2006,Gis2006,Yan2013}.
To prevent this kind of attack, a photon number splitter (or a photon
beam splitter (PBS): 50/50 \cite{Yan2015}) device and a wavelength
filter device could be adopted \cite{Den2005_4,Den2005_2,Li2006,Den2006_1}.

\section*{Acknowledgment}

We would like to thank the Ministry of Science and Technology of Republic
of China for financial support of this research under Contract No.
MOST 103-2221-E-006 -177 -.

\bibliographystyle{IEEEtran}
\addcontentsline{toc}{section}{\refname}\bibliography{ISLAB}

\begin{thebibliography}{10}
\providecommand{\url}[1]{#1}
\csname url@samestyle\endcsname
\providecommand{\newblock}{\relax}
\providecommand{\bibinfo}[2]{#2}
\providecommand{\BIBentrySTDinterwordspacing}{\spaceskip=0pt\relax}
\providecommand{\BIBentryALTinterwordstretchfactor}{4}
\providecommand{\BIBentryALTinterwordspacing}{\spaceskip=\fontdimen2\font plus
\BIBentryALTinterwordstretchfactor\fontdimen3\font minus
  \fontdimen4\font\relax}
\providecommand{\BIBforeignlanguage}[2]{{%
\expandafter\ifx\csname l@#1\endcsname\relax
\typeout{** WARNING: IEEEtran.bst: No hyphenation pattern has been}%
\typeout{** loaded for the language `#1'. Using the pattern for}%
\typeout{** the default language instead.}%
\else
\language=\csname l@#1\endcsname
\fi
#2}}
\providecommand{\BIBdecl}{\relax}
\BIBdecl

\bibitem{Yu2014}
K.-F. Yu, C.-W. Yang, C.-H. Liao, and T.~Hwang, ``Authenticated semi-quantum
  key distribution protocol using bell states,'' \emph{Quantum Information
  Processing}, vol.~13, no.~6, pp. 1457--1465, 2014.

\bibitem{Boy2007}
M.~Boyer, D.~Kenigsberg, and T.~Mor, ``Quantum key distribution with classical
  bob,'' \emph{Physical Review Letter}, vol.~99, p. 140501, 2007.

\bibitem{Boy2009}
M.~Boyer, R.~Gelles, D.~Kenigsberg, and T.~Mor, ``Semiquantum key
  distribution,'' \emph{Physical Review A}, vol.~79, no.~3, p. 032341, 2009.

\bibitem{Den2003}
F.~G. Deng, G.~L. Long, and X.~S. Liu, ``Two-step quantum direct communication
  protocol using the {Einstein-Podolsky-Rosen} pair block,'' \emph{Physical
  Review A}, vol.~68, p. 042317, 2003.

\bibitem{FIPS1995}
FIPS180-1, ``Secure hash standard,'' NIST, US Department of Commerce,
  Washington D. C., April 1995.

\bibitem{Pre1997}
B.~Preneel, H.~Dobbertin, and A.~Bosselaers, ``The cryptographic hash function
  {RIPEMD-160},'' \emph{Crypto Bytes}, vol.~3, no.~2, pp. 9--14, 1997.

\bibitem{Cal1996}
A.~R. Calderbank and P.~W. Shor, ``Good quantum error-correcting codes exist,''
  \emph{Physical Review A}, vol.~54, no.~2, pp. 1098--1105, 1996.

\bibitem{Cal1997}
A.~R. Calderbank, E.~M. Rains, P.~W. Shor, and N.~J.~A. Sloane, ``Quantum error
  correction and orthogonal geometry,'' \emph{Physical Review Letters},
  vol.~78, no.~3, pp. 405--408, 1997.

\bibitem{Eke1996}
A.~Ekert and C.~Macchiavello, ``Quantum error correction for communication,''
  \emph{Physical Review Letters}, vol.~77, no.~12, pp. 2585--2588, 1996.

\bibitem{Got1996}
D.~Gottesman, ``Class of quantum error-correcting codes saturating the quantum
  hamming bound,'' \emph{Physical Review A}, vol.~54, no.~3, pp. 1862--1868,
  1996.

\bibitem{Kni2000}
E.~Knill, R.~Laflamme, and L.~Viola, ``Theory of quantum error correction for
  general noise,'' \emph{Physical Review Letters}, vol.~84, no.~11, pp.
  2525--2528, 2000.

\bibitem{Laf1996}
R.~Laflamme, C.~Miquel, J.~P. Paz, and W.~H. Zurek, ``Perfect quantum error
  correcting code,'' \emph{Physical Review Letters}, vol.~77, no.~1, pp.
  198--201, 1996.

\bibitem{Sho1995}
P.~W. Shor, ``Scheme for reducing decoherence in quantum computer memory,''
  \emph{Physical Review A}, vol.~52, no.~4, pp. R2493--R2496, 1995.

\bibitem{Yan2012}
C.-W. Yang and T.~Hwang, ``Improved {QSDC} protocol over a collective-dephasing
  noise channel,'' \emph{International Journal of Theoretical Physics},
  vol.~51, no.~12, pp. 3941--3950, 2012.

\bibitem{Yan2013_3}
C.-W. Yang, T.~Hwang, and T.-H. Lin, ``Modification attack on qsdc with
  authentication and the improvement,'' \emph{International Journal of
  Theoretical Physics}, vol.~52, no.~7, pp. 2230--2234, 2013.

\bibitem{Hwa2014}
T.~Hwang, Y.-P. Luo, C.-W. Yang, and T.-H. Lin, ``Quantum authencryption:
  one-step authenticated quantum secure direct communications for off-line
  communicants,'' \emph{Quantum Information Processing}, vol.~13, no.~4, pp.
  925--933, 2014.

\bibitem{Cai2006}
Q.~Y. Cai, ``Eavesdropping on the two-way quantum communication protocols with
  invisible photons,'' \emph{Physics Letters A}, vol. 351, pp. 23--25, 2006.

\bibitem{Gis2006}
N.~Gisin, S.~Fasel, B.~Kraus, H.~Zbinden, and G.~Ribordy, ``Trojan-horse
  attacks on quantum-key-distribution systems,'' \emph{Physical Review A},
  vol.~73, no.~2, p. 022320, 2006.

\bibitem{Yan2013}
C.~W. Yang, T.~Hwang, and Y.~P. Luo, ``Enhancement on ''{Q}uantum blind
  signature based on two-state vector formalism'','' \emph{Quantum Information
  Processing}, vol.~12, no.~1, pp. 109--117, 2013.

\bibitem{Yan2015}
Y.-G. Yang, S.-J. Sun, and Q.-Q. Zhao, ``Trojan-horse attacks on quantum key
  distribution with classical bob,'' \emph{Quantum Information Processing},
  vol.~14, no.~2, pp. 681--686, 2015.

\bibitem{Den2005_4}
F.~G. Deng, P.~Zhou, X.~H. Li, C.~Y. Li, and H.~Y. Zhou, ``Robustness of
  two-way quantum communication protocols against {T}rojan horse attack,''
  \emph{arXiv:quant-ph/0508168v1}, 2005.

\bibitem{Den2005_2}
F.~G. Deng, X.~H. Li, H.~Y. Zhou, and Z.~J. Zhang, ``Improving the security of
  multiparty quantum secret sharing against {Trojan} horse attack,''
  \emph{Physical Review A}, vol.~72, no.~4, p. 044302, 2005.

\bibitem{Li2006}
X.~H. Li, F.~G. Deng, and H.~Y. Zhou, ``Improving the security of secure direct
  communication based on the secret transmitting order of particles,''
  \emph{Physical Review A}, vol.~74, no.~5, pp. 054\,302--1--054\,302--4, 2006.

\bibitem{Den2006_1}
F.~G. Deng, X.~H. Li, H.~Y. Zhou, and Z.~J. Zhang, ``Erratum: Improving the
  security of multiparty quantum secret sharing against {T}rojan horse attack
  [phys. rev. a 72, 044302 (2005)],'' \emph{Physical Review A}, vol.~73, no.~4,
  p. 049901, 2006.

\end{thebibliography}

\end{document}